\documentclass[12pt]{article}
\usepackage{amsmath,amsfonts}

\newcommand\maps{\colon}
\newcommand\ttto{{-}\!{-}\!{-}\!{\longrightarrow}}
\newcommand\e{{1\!\!\rm I}}
\renewcommand\H{{\mathbb H}}
\newcommand\R{{\mathbb R}}
\newcommand\Cl{{\mathfrak U}}
\newcommand\mbi[1]{{\text{\textbf{\textit{#1}}}}}
\newcommand\nn{\emptyset}
\newcommand\iCl{{\mathfrak U}^\star}
\newcommand\iso{\cong}
\newcommand\p{\gamma}

\title{\normalsize\bf Comment on ``Quantization of Diffeomorphism-Invariant 
      Theories with Fermions''}
\author{\em Alexander \t Iu.\ Vlasov \thanks{
       E-mail: Alexander.Vlasov@PObox.spbu.ru,
       alex@protection.spb.su}}
\date{}

\begin{document}

\maketitle

\begin{abstract}
 In the comment to the article by J.\ Baez and K.\ Krasnov (hep--th/9703112)
are discussed some topics related with application of certain constructions
to non-trivial principal bundles.
\end{abstract}

\noindent{---}\\
In chapter 2 of the article \cite{BK} is used following construction:

\begin{quote}
{\footnotesize (...)
First, define a `transporter' from the point $p$ to the
point $q$ to be a map from $P_p$ to $P_q$ that commutes with
the right action of $G$ on the bundle $P$.
If we trivialize the bundle over $p$ and $q$, we can think of
such a transporter simply as an element of $G$.
A `generalized connection' $A$ is a map assigning to
each oriented analytic path $e$ in $\Sigma$ a parallel transporter
$A_e \maps P_p \to P_q$, where $p$ is the initial point of the
path $e$ and $q$ is the final point.   We require that $A$ satisfy
certain obvious consistency conditions: $A$ should assign the
same transporter to two paths that differ only by an orientation-preserving
reparametrization, it should assign to the inverse of any path the
inverse transporter, and it should assign to the composite of
two paths the composite transporter.

An ordinary smooth connection $A$ gives a generalized connection where
the parallel transporter $A_e$ along any path $e$ is simply the holonomy of
$A$ along this path (...) 
}
\end{quote}

 The {\em trivialization} here is identification of fibers $P_p$
and $P_q$ with $G$ (structure group) in the both points $p$ and $q$, so
the `transporter' can be represented as left action of $G$ on $G$.

\begin{equation}
\begin{array}{ccccc}
P_p & \multicolumn{3}{c}{\ttto} & P_q \\
 & \searrow & & \swarrow & \\
 & & G & &
\end{array}
\label{triv}
\end{equation}

 Let us consider a connection on the principal bundle $P(M,G)$ with total
space $P$, structure group $G$ and base space $M$. $P_x$ is fiber in $x \in M$.
It should be emphasized that for open path it is necessary to consider an
additional structure, a trivialization at endpoints, while for closed path
there is a way to introduce holonomy group without using a trivialization.
For closed path the holonomy group has well defined ``internal'' description
as a group of automorphisms of a fiber and a choice of a point in the fiber
produces unique homomorphism to the group $G$ \cite{KN}. For open path $\p$
the parallel transport along given path also produces unique image of a point
$u_p \in P_p$, it is ending point $u_q \in P_q$ of horizontal lift of $\p$ with
initial point $u_q$, but it is the points in {\em different} fibers. To find an
element of $G$ for the point $u_p$ above $p$ and the point $u_q$ above $q$ it
is necessary to map both fibers to group $G$ as in diagram~(\ref{triv}).%
\footnote{There is other method for a given map $t_{p,q} \maps P_p \to P_q$.
Then it is possible to work with elements $m_q$ and $t_{p,q}(m_p)$ of
the same fiber. It will be discussed further.}

 It is always possible to find maps (\ref{triv}) for two given points of base
space $M$, but if it is necessary to consider {\em any path} on $M$, the
points $p$ and $q$ can be any points of $M$.

Is it possible to identify a fiber $P_x$ in any point $x \in M$ with group $G$?
There is simple proof that such {\em construction maybe continuous only for
trivial principal bundle}: let us identify fiber $P_x$ with group $G$ in
each point of $M$. Then for each point of $M$ it is possible to choose
point of $P_x$ that corresponds to unit {$(\e_G)_x$} of group $G$.
It produces {\em section} of principal bundle, but only trivial principal
bundles $P = M \times G$ can have global section \cite{KN,NS}.
For non-trivial principal bundle the fiber is equivalent with $G$ only as
with topological space without given structure of a group.

 Let us consider the trivial principal bundle. It is possible to consider
any fiber as group $G$ and the global trivialization produces a general
method to calculation of cylinder function \cite{BK}
\[   \Psi(A) = \psi({\cal P} \exp \int_{\p_1} A, \dots,
 {\cal P} \exp \int_{\p_n} A) \]
for analytic paths $\p_1,\dots,\p_n$ in $M$ with arbitrary endpoints. The
example is mentioned here, because only using of the trivialization, an
auxiliary structure in the model, make possible to calculate value of the
integrals above.
Let us compare the construction with some other gauge theories. In the
theories together with principal bundle of gauge field there is an associated
bundle, for example matter fields and any open path defines action of group
$G$ on section $F$ of the associated bundle $E(M,F,G,P)$.

The comparison justify consideration of trivialization as some analogue of
the additional structure for more straightforward work with integral above.
It is possible, because any trivial principal bundle defines unique
{\em canonical flat connection} with zero curvature \cite{KN}. The second
connection make possible to consider element $g \in G$ by comparison of two
horizontal lifts of the same path by two different connections.

The construction with two connections also works for non-trivial principal
bundle. Second connection here could not be treated as some trivialization,
but it make possible to calculate element of $G$ for any open path $\p$ and
given initial point $u \in P$. Instead of two different connections on the
same principal bundle it is possible to consider a second connection on
associated bundle $E(M,F{=}G,G,P)$ with a fiber is the same group $G$. Here
we have two different bundles with the same structure group, and it make the
construction similar to ``traditional'' gauge theories discussed earlier.

The construction could be considered as some
abstract exercise without relation with article under consideration, but let
us consider concrete example of application of the method.

\medskip

The $SU(2)$ group of spin network appears as $Spin(3)$, --- double cover of
$SO(3)$ group of 3{\bf D} space rotations. Let us recall direct construction
of the spin group with using of Clifford algebra \cite{MP,GM}.

The Clifford algebra $\Cl_{0,3}$ is isomorphic \cite{ClTab}
with $\H \oplus \H$. Here $\H$ is algebra of quaternions and element
of $\Cl_{0,3}$ can be represented as algebra of matrices:

$$
 {\bf a \oplus b } \equiv
 \left[ \begin{array}{cc} {\bf a} & 0 \\ 0 & {\bf b} \end{array} \right] \quad
 {\bf a, b} \in \H
$$

Three generators of the algebra $\Cl_{0,3}$ may be chosen as
${\mbi{e}_1 = (-\mbi{i})\oplus\mbi{i}'}$,
${\mbi{e}_2 = (-\mbi{j})\oplus\mbi{j}'}$,
${\mbi{e}_3 = (-\mbi{k})\oplus\mbi{k}'}$
where $\mbi{i},\mbi{j},\mbi{k}$ and $\mbi{i}',\mbi{j}',\mbi{k}'$ are three
quaternionic units $\mbi{i}^2=\mbi{j}^2=\mbi{k}^2 = -1 $ of first and
second terms of direct sum. Then $\mbi{e}_n\mbi{e}_m = -\mbi{e}_m\mbi{e}_n$,
$m \neq n$, ${\mbi{e}_n}^2 = -{\bf 1} \equiv (-1) \oplus (-1)$, $m,n = 1,2,3$.

A group $\iCl_{0,3}$ of all invertible elements of $\Cl_{0,3}$ can be also
written with {\em multiplicative} notation as
$\H_{^\nn}\times\H_{^\nn}$, where $\H_{^\nn} \equiv \H -\{0\}$.
A $pin(3)$ is a group generated by multiplication of arbitrary number of
elements $\sum a_n\mbi{e}_n$, $\sum a_n^2 = 1$ and a $Spin(3)$ subgroup
is composed by only even number of such terms. So, the $Spin(3)$ group
could be written as $\sum c_\alpha e_\alpha$, $\sum c_\alpha^2 = 1$
where $e_\alpha$ are {\em four}
different even combinations of $\mbi{e}_n$, {\em i.e.}:
${e_0 \equiv {\bf 1} = -{\mbi{e}_n}^2 = 1 \oplus 1}$,
${e_1 \equiv -\mbi{e}_2\mbi{e}_3 = \mbi{i}\oplus\mbi{i}'}$,
${e_2 \equiv -\mbi{e}_3\mbi{e}_1 = \mbi{j}\oplus\mbi{j}'}$,
${e_3 \equiv -\mbi{e}_1\mbi{e}_2 = \mbi{j}\oplus\mbi{j}'}$.
The subgroup is isomorphic with $SU(2)$ and we have following structure
of the group $\iCl_{0,3}$ :
$\H_{^\nn}\times\H_{^\nn} = \R_+ \times SU(2) \times \R_+ \times SU(2)$,
where $\R_+$ is multiplicative group of positive real numbers.

\medskip

Let us now consider the Clifford algebra bundle $\Cl_{0,3}$ on a manifold. It
is possible \cite{GM}, but already existence of $\iCl_{0,3}$ bundles is not
guaranteed globally. Let us suppose for simplicity that there the $\iCl_{0,3}$
principal bundle is exist and look for conditions for $Spin(3)$ bundle.

 {\em A reduction of principle bundle with structure group $G$ to closed
subgroup $H$ is possible if and only if associated bundle with fiber $G/H$
accepts a global section} \cite{KN} . It is possible for particular case
of $G/H$ is diffeomorphic to Euclidean space \cite{KN} , like the subgroups
$\R_+\sim\R$, so the reduction of $\iCl_{0,3}$ to $SU(2) \times SU(2)$
subgroup is always possible.

 It is possible now to work with the $SU(2) \times SU(2) \iso Spin(4)$ group%
\footnote{The isomorphism with $Spin(4)$ make the consideration relevant
also to some theories with Euclidean gravity.} instead of $\iCl_{0,3}$ and to
consider a reduction of the group to $SU(2) \iso Spin(3)$. The condition
of the reduction is existance of a section of associated bundle with
fiber $G/H$, {\em i.e.} the same group $SU(2) \iso Spin(4)/Spin(3)$.

\smallskip

 {\em Summarizing the example }: If it is possible to make reduction of
$\iCl_{0,3}$ multiplicative group of Clifford bundle on manifold if and
only if associated bundle with fiber $SU(2)$ accepts a global section and
if it is possible we have two different bundles with the same structure
group $SU(2)$: $Spin(3)$ subgroup of initial $\iCl_{0,3}$ (or $Spin(4)$)
bundle and quotient group $SU(2) \iso Spin(4)/SU(2)$ of associated bundle.
The associated bundle has global section, but principal $SU(2)$ bundle,
reduction of initial $\iCl_{0,3}$ or $Spin(4)$ bundle may be either trivial
or not, {\em i.e.} it may have no global section.

\medskip

\noindent{\bf Conclusion} The example has certain analogies with some
properties of construction was used in model was discussed earlier.


\begin{thebibliography}{99}

\bibitem{BK} J. C. Baez, K. V. Krasnov, Quantization of
 Diffeomorphism-Invariant Theories with Fermions; hep-th/9703112.


\bibitem{KN} S. Kobayashi, K. Nomizu, {\em Foundations of Differential
 Geometry}, Vol. 1, Interscience Publ., 1963.

\bibitem{NS} N. Steenrod, {\em Topology of Fibre Bundles}, Princeton
 Univ. Press, 1951.

\bibitem{MP} M. M. Postnikov, {\em Lie Algebras and Groups}, Moscow,
 Nauka, 1982. [Russ. {\em Gruppy i Algebry Li }]

\bibitem{GM} J. E. Gilbert, M. A. M. Murray, {\em Clifford algebras and
 Dirac operators in harmonic analysis}, Cambridge Univ. Press, 1991.

\bibitem{ClTab} I. R. Porteous, {\em Clifford Algebras and Classical
 Groups}, Cambridge Univ. Press, 1995.


\end{thebibliography}
\end{document}